\documentclass[aps,prb,superscriptaddress,amsfonts,amsmath,amssymb,reprint,showpacs,floatfix]{revtex4-1}

\usepackage{url}
\usepackage{bm}
\usepackage{graphicx}
\usepackage{amsmath}
\usepackage{amstext}
\usepackage{amssymb}
\usepackage{amsfonts}
\usepackage{amsbsy}
\usepackage{verbatim}
\usepackage{color}
\usepackage[colorlinks=true, urlcolor=blue, linkcolor=blue, citecolor=blue, pdftex]{hyperref}
\usepackage{multirow}
\usepackage[final]{pdfpages}
\usepackage{floatrow}
\begin{document}

\title{Gapless spin-liquid phase in the kagome spin-$\frac{1}{2}$ Heisenberg antiferromagnet}

\author{\href{http://www.ictp.it/research/cmsp/members/postdoctoral-fellows/yasir-iqbal.aspx}{Yasir Iqbal}}
\email[]{yiqbal@ictp.it}
\affiliation{The Abdus Salam International Centre for Theoretical Physics, P.O.~Box 586, I-34151 Trieste, Italy}
\affiliation{Laboratoire de Physique Th\'eorique UMR-5152, CNRS and 
Universit\'e de Toulouse, F-31062 Toulouse, France}
\author{\href{http://people.sissa.it/~becca/}{Federico Becca}}
\email[]{becca@sissa.it}
\affiliation{Democritos National Simulation Center, Istituto
Officina dei Materiali del CNR and SISSA$-$International School for Advanced Studies, Via Bonomea 265, I-34136 Trieste, Italy}
\author{\href{http://people.sissa.it/~sorella/}{Sandro Sorella}}
\email[]{sorella@sissa.it}
\affiliation{Democritos National Simulation Center, Istituto
Officina dei Materiali del CNR and SISSA$-$International School for Advanced Studies, Via Bonomea 265, I-34136 Trieste, Italy}
\author{\href{http://www.lpt.ups-tlse.fr/spip.php?article32}{Didier Poilblanc}}
\email[]{didier.poilblanc@irsamc.ups-tlse.fr}
\affiliation{Laboratoire de Physique Th\'eorique UMR-5152, CNRS and 
Universit\'e de Toulouse, F-31062 Toulouse, France}

\date{\today}

\begin{abstract}
We study the energy and the static spin structure factor of the ground state
of the spin-$1/2$ quantum Heisenberg antiferromagnetic model on the kagome 
lattice. By the iterative application of a few Lanczos steps on accurate projected
fermionic wave functions and the Green's function Monte Carlo technique, 
we find that a gapless (algebraic) $U(1)$ Dirac spin liquid is competitive 
with previously proposed gapped (topological) $\mathbb{Z}_{2}$ spin liquids. By performing a 
finite-size extrapolation of the ground-state energy, we obtain an energy per
site $E/J=-0.4365(2)$, which is equal, within three error bars, to the estimates
given by the density-matrix renormalization group (DMRG). Our estimate is obtained
for a translationally invariant system, and, therefore, does not suffer from 
boundary effects, like in DMRG. Moreover, on finite toric clusters at the pure variational level,
our energies are lower compared to those from DMRG calculations.
\end{abstract}

\pacs{75.10.Jm, 75.10.Kt, 75.40.Mg, 75.50.Ee}

\maketitle

{\it Introduction}.
The spin-$1/2$ quantum Heisenberg antiferromagnet (QHAF) on the kagome lattice
provides a conducive environment to stabilize a quantum paramagnetic phase of 
matter down to zero temperature,~\cite{Anderson-1973,Anderson-1987,Balents-2010}
a fact that has been convincingly established theoretically from several 
studies, including exact diagonalization,~\cite{Elser-1989,Lecheminant-1997,Sindzingre-2009,Nakano-2011,Lauchli-2011} 
series expansion,~\cite{Singh-1992,Misguich-2005} quantum Monte 
Carlo,~\cite{Dang-2011} and analytical techniques.~\cite{Balents-2002} 
The question of the precise nature of the spin-liquid state of the kagome 
spin-$1/2$ QHAF has been intensely debated on the theoretical front, albeit 
without any definitive conclusions. Different approximate numerical techniques
have claimed a variety of ground states. On the one hand, density-matrix 
renormalization group (DMRG) calculations have been claimed for a fully gapped 
(nonchiral) $\mathbb{Z}_{2}$ topological spin-liquid ground state that does 
not break any point group symmetry.~\cite{Yan-2011,Depenbrock-2012} 
On the other hand, an algebraic and fully symmetric $U(1)$ Dirac spin liquid 
has been proposed as the ground state, by using projected fermionic wave 
functions and the variational Monte Carlo (VMC) 
approach.~\cite{Ran-2007,Hermele-2008,Ma-2008,Iqbal-2011a,Iqbal-2011b,Iqbal-2012} 
In addition, valence bond crystals have been suggested from many other 
techniques. In particular, a $36$-site unit cell valence-bond 
crystal~\cite{Marston-1991,Nikolic-2003,Hwang-2011} was proposed using quantum
dimer models,~\cite{Zeng-1995,Poilblanc-2010,Schwandt-2010,Ralko-2010,Poilblanc-2011} 
series expansion~\cite{Singh-2007,Singh-2008} and multiscale entanglement 
renormalization ansatz (MERA)~\cite{Evenbly-2010} techniques. Finally, a
recent coupled cluster method (CCM) suggested a $q=0$ (uniform) 
state.~\cite{Gotze-2011}

On general theoretical grounds, the $\mathbb{Z}_{2}$ spin liquids in two
spatial dimensions are known to be stable 
phases,~\cite{Sachdev-1992,Wen-2002,Misguich-2002}
as compared to algebraic $U(1)$ spin liquids, which are known to be only 
marginally stable.~\cite{Hermele-2004} However, explicit numerical calculations
using projected wave functions have shown the $U(1)$ Dirac spin liquid to be 
stable (locally and globally) with respect to dimerizing into all known 
valence-bond crystal phases.~\cite{Ran-2007,Ma-2008,Iqbal-2011a,Iqbal-2012} 
Furthermore, it was shown that, within this class of Gutzwiller projected wave
functions, all the fully symmetric, gapped $\mathbb{Z}_{2}$ spin liquids have
a higher energy compared to the $U(1)$ Dirac spin 
liquid.~\cite{Iqbal-2011b,Yang-2012}

On the experimental front, the kagome spin-$1/2$ QHAF model is well reproduced
in herbertsmithite [ZnCu$_{3}$(OH)$_{6}$Cl$_{2}$], a compound with perfect 
kagome lattice geometry.~\cite{Shores-2005,Ofer-2006,Bert-2007,Lee-2007,Lee-2008,Vries-2008,Imai-2008,Olariu-2008,Mendels-2010,Han-2011} 
All experimental probes on herbertsmithite point towards a spin-liquid behavior
down to $20$ mK (i.e., four orders of magnitude smaller than the superexchange
coupling), which was established on the magnesium version of herbertsmithite 
[MgCu$_{3}$(OH)$_{6}$Cl$_{2}$].~\cite{Mendels-2007,Helton-2007,Kermarrec-2011}
Raman spectroscopic studies give further hints towards a gapless (algebraic) 
spin liquid.~\cite{Wulferding-2010}

In this Rapid Communication, we systematically improve the projected fermionic wave 
functions of the $U(1)$ Dirac and other competing spin liquids by applying a 
few Lanczos steps on large clusters, implemented stochastically within a 
variational Monte Carlo method.~\cite{Sorella-2001} We perform a zero-variance 
extrapolation of the energy and the static spin structure factor, which enables
us to extract their exact values in the ground state on large cluster sizes 
and obtain an accurate estimate of the thermodynamic limit. In addition, 
we use the Green's function Monte Carlo method, with the fixed-node (FN) 
approximation,~\cite{Haaf-1995} to extract the physical properties of the 
true ground state. Our main result is to show that the $U(1)$ gapless spin 
liquid has an energy quite close to recent DMRG 
estimates,~\cite{Yan-2011,Depenbrock-2012} thus representing a very competitive
state for the spin-$1/2$ QHAF on the kagome lattice (if not the true ground 
state).

\begin{table*}
\centering
\begin{tabular}{lllllllcl}
 \hline \hline
       \multicolumn{1}{c}{Size}
    & \multicolumn{1}{c}{$0$-LS}
    & \multicolumn{1}{c}{$1$-LS} 
    & \multicolumn{1}{c}{$2$-LS}
    & \multicolumn{1}{c}{$0$-LS $+$ FN}
    & \multicolumn{1}{c}{$1$-LS $+$ FN}
    & \multicolumn{1}{c}{$2$-LS $+$ FN}
    & \multicolumn{1}{c}{Var. DMRG}  
    & \multicolumn{1}{c}{Est. ground state}  \\ \hline
       
\multirow{1}{*}{$48$} & $-0.4293510(4)$ & $-0.4352562(3)$ & $-0.436712(1)$ & $-0.432130(2)$ & $-0.435834(3)$ & $-0.436942(2)$ & $-0.4366$ &$\bm{-0.437845(4)}$  \\ 
                                                                                                                       
\multirow{1}{*}{$108$} & $-0.4287665(4)$ & $-0.4341032(5)$ & $-0.435787(3)$ & $-0.431507(1)$ & $-0.434823(2)$ & $-0.436072(1)$ & $-0.4316$ & $\bm{-0.437178(9)}$  \\ 
                                                                                    
\multirow{1}{*}{$144$} & $-0.4286959(5)$ & $-0.4337616(4)$ & $-0.435515(4)$ & $-0.4314455(8)$ & $-0.434544(2)$ & $-0.435839(9)$ &  & $\bm{-0.43698(2)}$ \\ 
                                                      
\multirow{1}{*}{$192$} & $-0.4286749(4)$ & $-0.4334481(5)$ & $-0.435255(4)$ & $-0.431437(2)$ & $-0.434325(4)$ & $-0.435633(8)$ &  & $\bm{-0.43674(3)}$  \\ \hline \hline

\end{tabular}
\caption{\raggedright Energies of the $U(1)$ Dirac spin liquid with $p=0$, 
$1$, and $2$ Lanczos steps on different cluster sizes obtained by variational 
and FN Monte Carlo are given. In the penultimate column, we report the best 
{\it variational} DMRG energies (Ref.~\onlinecite{privateDMRG}). The ground-state energy of
the spin-$1/2$ QHAF estimated by us using zero-variance extrapolation of VMC 
energy values on different cluster sizes is marked in bold.}
\label{tab:en-lanczos}
\end{table*}

{\it Model, wave functions, and numerical techniques}.

The Hamiltonian for the spin-$1/2$ quantum Heisenberg antiferromagnetic model 
is
\begin{equation}
\label{eqn:heis-ham}
\hat{{\cal H}} = J \sum_{\langle ij \rangle} \mathbf{\hat{S}}_{i} \cdot \mathbf{\hat{S}}_{j},
\end{equation} 
where $J>0$ and $\langle ij \rangle$  denotes the sum over nearest-neighbor pairs 
of sites. The $\mathbf{\hat{S}}_{i}$ are spin-$1/2$ operators at each site $i$.
All energies will be given in units of $J$.

The physical variational wave functions are defined by projecting 
noncorrelated fermionic states:
\begin{equation}
\label{eqn:var-wf}
|\Psi_{{\rm VMC}}(\chi_{ij},\Delta_{ij},\mu,\zeta)\rangle=\mathbf{{\cal P}_{G}}|\Psi_{{\rm MF}}(\chi_{ij},\Delta_{ij},\mu,\zeta)\rangle,
\end{equation}
where $\mathbf{{\cal P}_{G}}=\prod_{i}(1-n_{i,\uparrow}n_{i,\downarrow})$ is
the full Gutzwiller projector enforcing the one fermion per site constraint.
Here, $|\Psi_{{\rm MF}}(\chi_{ij},\Delta_{ij},\mu,\zeta)\rangle$ is the ground
state of a mean-field Hamiltonian constructed out of Abrikosov fermions and 
containing hopping, chemical potential, and singlet pairing terms:
\begin{eqnarray}
\label{eqn:MF0}
\hat{{\cal H}}_{{\rm MF}} &=& \sum_{i,j,\alpha} (\chi_{ij}+\mu\delta_{ij})\hat{c}_{i,\alpha}^{\dagger}\hat{c}_{j,\alpha} \nonumber \\
&+&\sum_{i,j} \{(\Delta_{ij}+\zeta\delta_{ij})\hat{c}^{\dagger}_{i,\uparrow}\hat{c}^{\dagger}_{j,\downarrow}+{\rm H.c.}\} \, ,
\end{eqnarray}
where $\alpha=\uparrow,\downarrow$, $\chi_{ij}=\chi_{ji}^{*}$, and $\Delta_{ij}=\Delta_{ji}$. 
Besides the chemical potential $\mu$, we also consider real and imaginary 
components of on-site pairing, which are absorbed in $\zeta$.
The spin-liquid phases are characterized by different patterns of distribution
of the underlying gauge fluxes through the plaquettes which are implemented 
by a certain distribution of the phases of $\chi_{ij}$ and $\Delta_{ij}$ on the
lattice links; in addition one also needs to specify the on-site terms 
$\mu$ and $\zeta$.~\cite{Wen-1991,Wen-2002}

Here, we want to improve previous variational calculations, and approach the true ground state in a systematic way. This task can be achieved by the
application of few Lanczos steps:~\cite{Sorella-2001}
\begin{equation}
\label{eqn:psi-ls}
|\Psi_{p-\rm{LS}}\rangle =  \bigg{(}1+\sum_{k=1}^{p}\alpha_{k}\hat{{\cal H}}^{k}\bigg{)}|\Psi_{\rm VMC}\rangle,
\end{equation}
where the $\alpha_{k}$'s are additional variational parameters. The convergence
of $|\Psi_{p\text{-}\rm{LS}}\rangle$ to the exact ground state $|\Psi_{\rm ex}\rangle$ 
is guaranteed for large $p$ provided the starting state is not orthogonal to
$|\Psi_{\rm ex}\rangle$, i.e., 
for $\langle\Psi_{\rm ex}|\Psi_{\rm VMC}\rangle \neq 0$. 
However, on large cluster sizes, only a few steps can be efficiently performed
and here we consider the case with $p=1$ and $p=2$ ($p=0$ corresponds to the 
original starting variational wave function). Subsequently, an estimate of 
the exact ground-state energy may be achieved by the method of variance 
extrapolation: For sufficiently accurate states, we have that 
$E\approx E_{\rm ex}+{\rm constant}\times\sigma^{2}$, where 
$E=\langle\hat{{\cal H}}\rangle/N$ and 
$\sigma^{2}=(\langle\hat{{\cal H}}^{2}\rangle-\langle\hat{{\cal H}}\rangle^{2})/N$ 
are the energy and variance per site, respectively, whence, the exact 
ground-state energy $E_{\rm ex}$ can be extracted by fitting $E$ vs 
$\sigma^{2}$ for $p=0$, $1$, and $2$.

\begin{figure}[b]
\includegraphics[width=1.08\columnwidth]{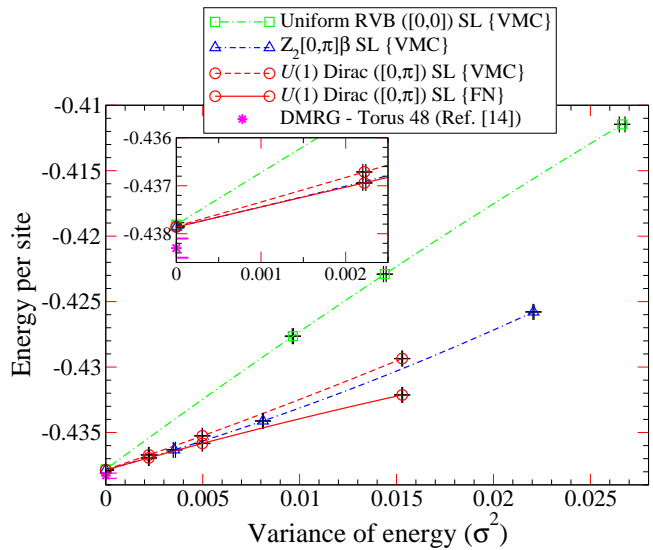}
\caption{\label{fig:ener48}
(Color online) Variational energies of the $48$-site cluster as a function
of the variance of the energy, for zero, one, and two Lanczos steps. The 
ground-state energy is estimated by extrapolating the three variational results
to the zero-variance limit by a quadratic fit. Three different starting wave 
functions are used. The $U(1)$ Dirac spin liquid has also been studied using 
FN approximation.}
\end{figure}

The energy, its variance, and other physical properties of the wave functions 
corresponding to $p=0$, $1$, and $2$ Lanczos steps are obtained using the standard
VMC method. Moreover, the pure variational approach may be improved by using
the FN approach, in which the high-energy components of the variational wave 
function are (partially) filtered out.~\cite{Haaf-1995} In particular, 
in the FN Monte Carlo method, the ground state of an auxiliary FN Hamiltonian
is obtained and the approximation consists in assigning the nodal surface 
{\it a priori}, based upon a given guiding wave function, which is generally 
the best variational state. The energies obtained in this way are 
variational,~\cite{Haaf-1995} and hence we have a controlled approximation of 
the original problem. Here, the guiding wave function is obtained by optimizing
the mean-field state of Eq.~(\ref{eqn:var-wf}) using the method described in 
Refs.~\onlinecite{Sorella-2005,Yunoki-2006}. Then, we find the best Lanczos 
parameters $\alpha_{p}$ and finally we perform the VMC and FN Monte Carlo 
calculations for $|\Psi_{p\text{-}{\rm LS}}\rangle$ with $p=0$, $1$, and $2$.

\begin{figure}
\vspace{-0.23cm}
\includegraphics[width=1.08\columnwidth]{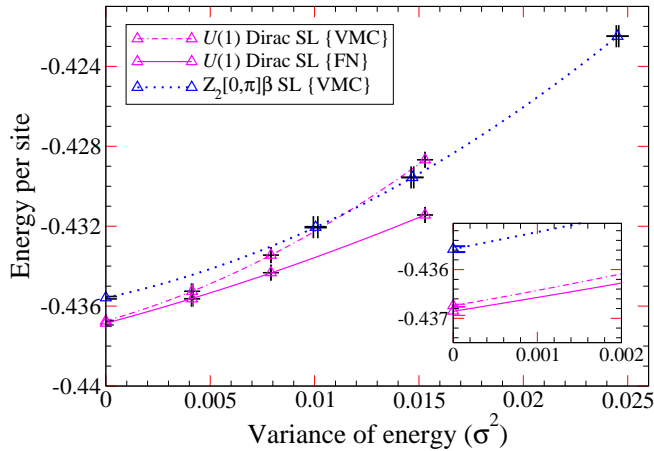}
\caption{\label{fig:enersizes}
(Color online) The same as in Fig.~\ref{fig:ener48} for the $192$-site cluster.
Here, only $\mathbb{Z}_{2}[0,\pi]\beta$ and $U(1)$ Dirac states have been
considered.}
\end{figure}

{\it Results}.
We performed our variational calculations on toric clusters with mixed 
periodic-antiperiodic boundary conditions on the mean-field Hamiltonian
of Eq.~(\ref{eqn:MF0}), which ensure nondegenerate wave functions at half 
filling. We first consider the $48$-site cluster (i.e., $4 \times 4 \times 3$).
As our starting ($p=0$) variational wave functions, we take three different 
spin liquids, namely, (i) the $U(1)$ Dirac spin liquid, which has a Fermi 
surface consisting of two points.~\cite{Ran-2007,Hermele-2008} The structure 
of the wave function is such that $10\%$ of the configurations $|x\rangle$ 
(in which electrons reside on different sites of the lattice with given spin 
along the $z$ direction) have zero weight (i.e., $\langle x|\Psi_{\rm VMC}\rangle=0$); 
(ii) the uniform RVB spin liquid, which consists of a large circular spinon 
Fermi surface,~\cite{Ma-2008} and has $35\%$ of the configurations with zero 
weight; and (iii) the $\mathbb{Z}_{2}[0,\pi]\beta$ spin liquid, which is fully 
gapped~\cite{Lu-2011} and has a negligible ($0.001\%$) number of configurations
with zero weight. The zero-weight configurations are not visited by the random
walk in the variational Monte Carlo method. The effect of two Lanczos steps 
on these wave functions is shown in Fig.~\ref{fig:ener48} [see also 
Table~\ref{tab:en-lanczos} for the actual values of the energies of the $U(1)$
Dirac state]. Our estimate of the ground-state energy on the $48$-site cluster
is thus $E/J=-0.437~845(4)$, which is comparable with the DMRG estimate on a 
torus.~\cite{Depenbrock-2012,Jiang-2008} Also the best {\it pure} variational 
energies are comparable within the two methods (see Table~\ref{tab:en-lanczos}).
We want to stress the fact that the extrapolated energy is the same (within 
error bars) upon starting from all three wave functions. This is mainly due 
to the fact that, on relatively small clusters, a few Lanczos steps are enough 
to filter out the high-energy components of the initial wave function and get 
a good estimation of the ground-state energy.

On larger sizes, the extrapolations of $U(1)$ and $\mathbb{Z}_{2}[0,\pi]\beta$
states deviate, the former one giving a slightly lower extrapolation (see 
Fig.~\ref{fig:enersizes} for the $192$-site cluster). This fact suggests that 
the actual ground state is better described by a gapless algebraic $U(1)$ 
Dirac state, rather than a gapped topological $\mathbb{Z}_{2}$ spin liquid, 
as reported by DMRG calculations. In the following, for obtaining the 
ground-state energies on larger clusters we used only the $U(1)$ Dirac wave 
function as the starting variational state. In Table~\ref{tab:en-lanczos}, 
we report our best results on different clusters (see the Supplemental 
Material\cite{suppmat} for plots of the variance extrapolations on $108$- and $144$-site clusters). We would like to emphasize that our best variational energy on a $108$-site cluster is significantly lower
compared to the corresponding DMRG one (see Table~\ref{tab:en-lanczos}). 

\begin{figure}
\includegraphics[width=1.05\columnwidth]{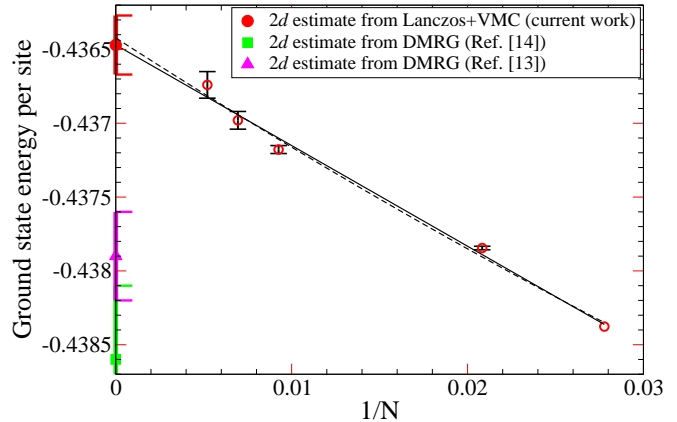}
\caption{\label{fig:enerthermo}
(Color online) The thermodynamic estimate of the ground-state energy obtained 
by a finite-size extrapolation of the estimated ground-state energies (see 
Table~\ref{tab:en-lanczos}). The linear (solid line) and quadratic (dashed 
line) fits give essentially the same estimate. The energy on the $36$-site 
cluster is from exact diagonalization. Comparison is also made with recent 
DMRG estimates.~\protect\cite{Yan-2011,Depenbrock-2012}}
\end{figure}

By using the ground-state energy estimates on different cluster sizes, we 
performed a finite-size extrapolation (see Fig.~\ref{fig:enerthermo}). 
Our final estimate for the energy of the infinite two-dimensional system is
\begin{equation}
E_{\infty}^{2D}/J=-0.4365(2).
\end{equation}
This estimate is slightly higher (see Fig.~\ref{fig:enerthermo}) compared to 
DMRG extrapolations of Refs.~\onlinecite{Yan-2011,Depenbrock-2012}. However, an 
energy estimate which is slightly lower by only a few error bars does not 
necessarily mean it is more accurate. We would like to stress that the same 
value for the extrapolated energy is obtained by using the FN approach (see the Supplemental Material\cite{suppmat}). Moreover, it is worth 
mentioning that our energies are obtained with a state that has all the 
symmetries of the lattice, while DMRG states are nonuniform (due to boundary 
effects).
 
Let us now move to the calculation of the spin-spin correlations, which is 
defined by
\begin{equation}
S({\bf q}) = \frac{1}{N} \sum_{ij}\sum_{\bf R} e^{-\imath {\bf q}\cdot {\bf R}} S_{ij}({\bf R}),
\end{equation} 
where $N$ is the total number of sites, $i,j=1$, $2$, and $3$ label the three 
sites in the unit cell, {\bf R} defines the Bravais lattice, and
$S_{ij}({\bf R})$ is the real space spin-spin correlation function. 

\begin{figure}
\includegraphics[width=1.0\columnwidth]{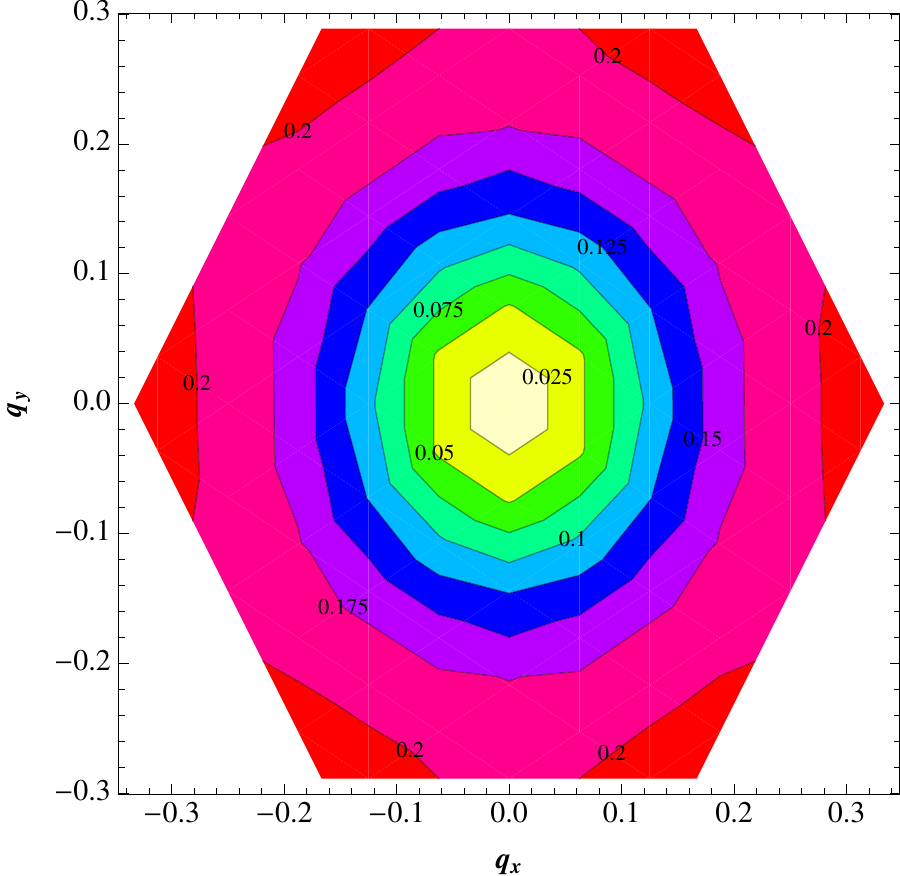}
\caption{\label{fig:sqmain}
(Color online) Intensity plot of the static spin structure factor $S$({\bf q}) 
on the $192$-site cluster.}
\end{figure}

The $U(1)$ Dirac spin liquid is characterized by a power-law ($\sim1/r^{4}$) 
decay of real-space, long-distance spin-spin correlations.~\cite{Hermele-2008}
Here, we study the evolution of its static spin structure factor $S$({\bf q}) 
on the $192$-site cluster under the action of one and two Lanczos steps and 
zero-variance extrapolation. Our estimate of the ground-state $S$({\bf q}) is 
obtained by a zero-variance extrapolation (see the Supplemental Material\cite{suppmat}). 
The corresponding intensity plot of the extrapolated $S$({\bf q}) is shown in 
Fig.~\ref{fig:sqmain}. One can clearly see that at large {\bf q}, the spectral
weight is concentrated on the corners of the hexagon, not very differently 
from what is found in a recent DMRG study.~\cite{Depenbrock-2012} 
However, what really matters is the behavior of $S$({\bf q}) for small {\bf q}
(namely, at long distance). Although the application of a few Lanczos steps may
not be sufficient to change the long-distance properties (because the 
Hamiltonian is a local operator), our calculations show that $S$({\bf q}) at
small {\bf q} remains practically unchanged under the action of one or two 
Lanczos steps and the subsequent zero-variance extrapolation (see the Supplemental Material\cite{suppmat}).

{\it Summary}.
In summary, our systematic numerical study shows that competitive variational
wave functions based upon Abrikosov fermions may be obtained. Indeed, our 
estimation for the energy of a gapless (algebraic) $U(1)$ Dirac spin liquid is
very close to the recent DMRG results,~\cite{Yan-2011,Depenbrock-2012} which 
supported a fully gapped $\mathbb{Z}_{2}$ topological spin-liquid ground state.
In this respect, our results lend support to the view that the exotic 
algebraic spin liquid can in fact occur as a true ground state of the 
spin-$1/2$ QHAF on the kagome lattice. Very recently, other approximate 
approaches proposed alternative ground states with or without broken 
symmetries.~\cite{Motrunich-2011a,Clark-2012,Capponi-2012} 

We would like to mention that a further improvement of the variational wave 
function would require an introduction of local monopole fluctuations over the
static mean-field state of Eq.~(\ref{eqn:MF0}). On small system sizes, such 
fluctuations were shown to lower the energy of the system within the Schwinger
boson approach.~\cite{Motrunich-2011} However, on large clusters, it is 
extremely difficult to construct workable wave functions with (even static) 
topological defects. It is worth mentioning that the possibility of another 
energetically competing state entering the game remains open; this is a chiral
$\mathbb{Z}_{2}$ topological spin liquid~\cite{Yang-1993} which has been 
proposed as the ground state within a Schwinger boson mean-field 
theory,~\cite{Messio-2011} but whose projected wave-function study remains to 
be done on large clusters such as $48$ sites so as to enable a comparison 
with the $U(1)$ Dirac spin liquid. Finally, the projected wave functions can 
also be constructed for chiral valence-bond crystal phases and it would be 
interesting to study their energetics, especially in light of the fact that 
they have been proposed as a competing ground state using generalized quantum 
dimer models.~\cite{Poilblanc-2011}

{\it Acknowledgments}
Y.I. and D.P. acknowledge support from the ``Agence Nationale de la Recherche''
under Grant No.~ANR~2010~BLANC~0406-0. We are grateful for the permission 
granted to access the HPC resources of CALMIP under the allocation 2012-P1231.

\newpage



\section*{Supplementary material}

# Supplementary Material

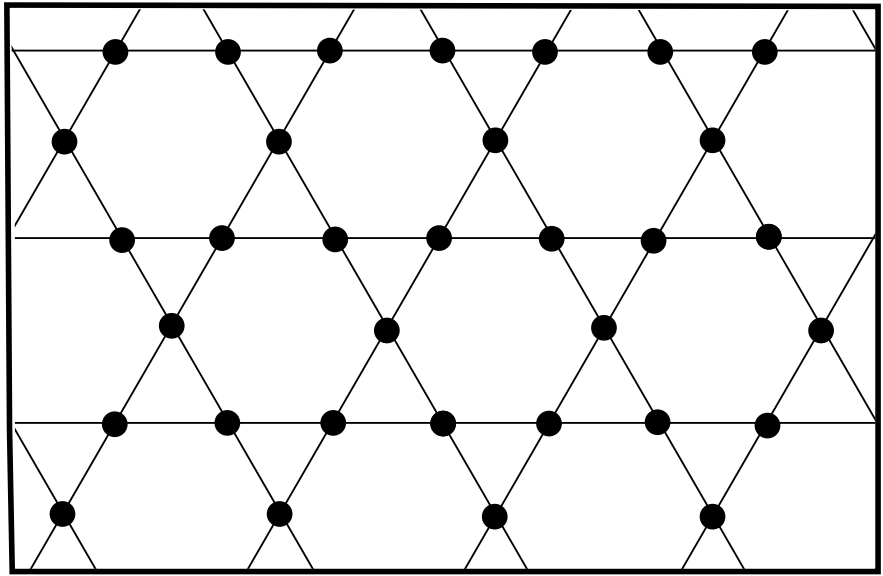

FIG. 5. A lattice of corner sharing triangles giving rise to two elementary plaquettes, the triangle and the hexagon. The kagome lattice is the most frustrated among the 11 Archimedean tilings possible in $2D$ and has a coordination number of 4.

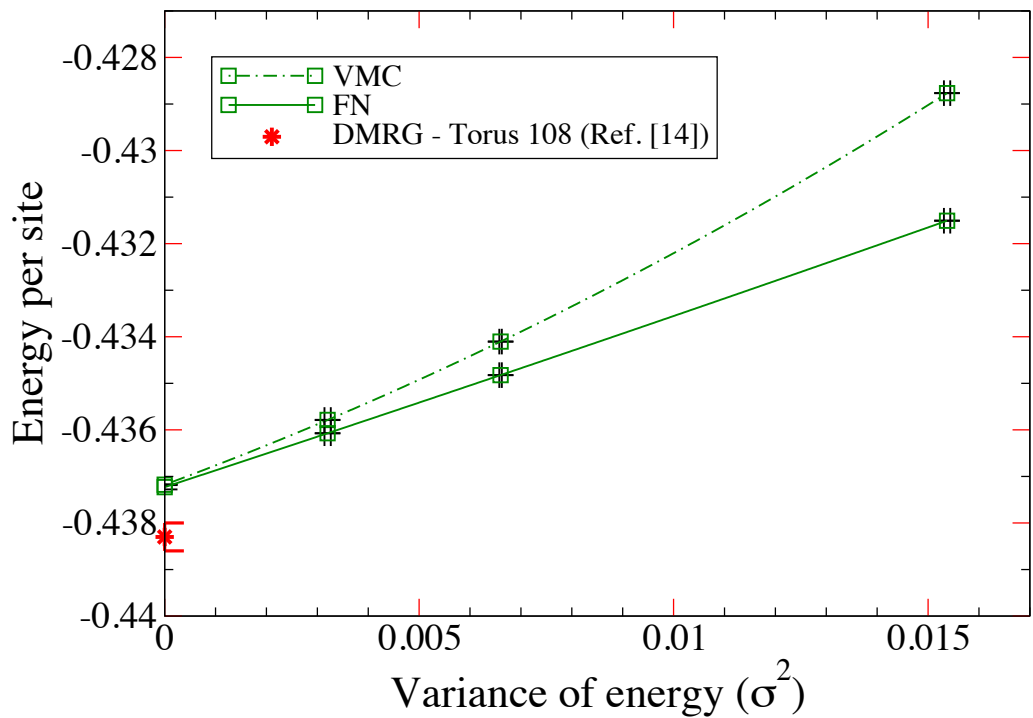

FIG. 6. (Color online) The same as in Fig. 1 of the main paper, for the 108-site cluster. Here, only the $U(1)$ Dirac state is considered and has been studied using both variational and fixed-node Monte Carlo methods.

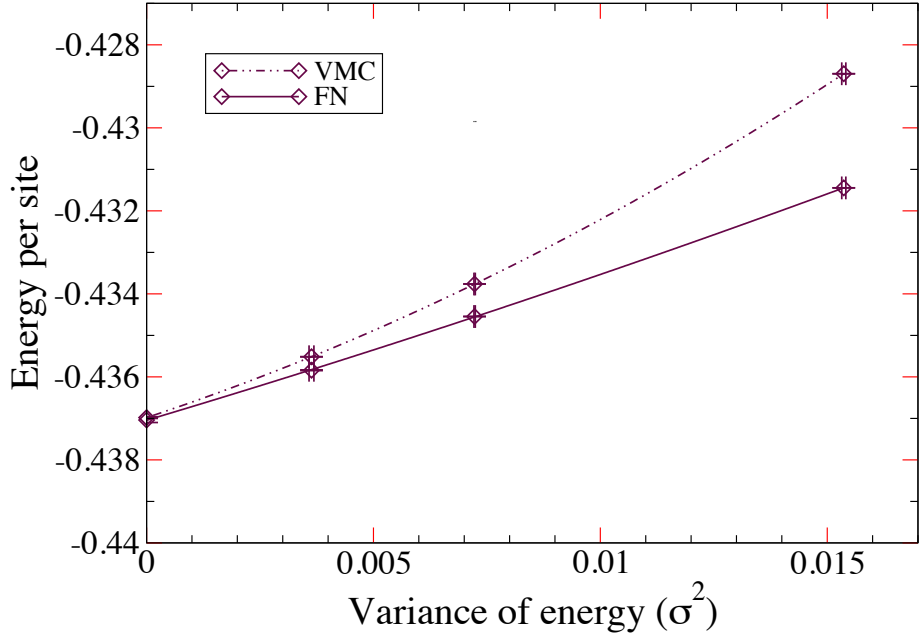

FIG. 7. (Color online) The same as in Fig. 1 of the main paper, for the 144-site cluster. Here, only the $U(1)$ Dirac state is considered and has been studied using both variational and fixed-node Monte Carlo methods.

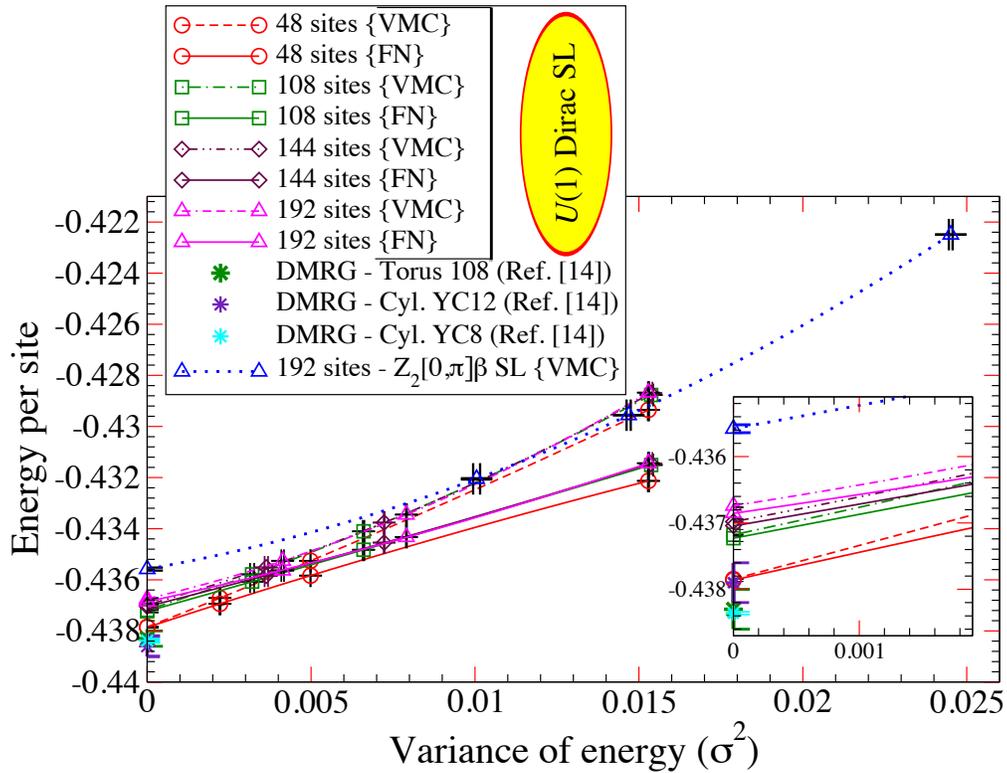

FIG. 8. (Color online) The variance extrapolations on all cluster sizes for different starting variational wave functions are shown together for comparison. The inset shows the magnification near zero-variance. Note: The magnitude of the pairing in the $\mathbb{Z}_2[0,\pi]\beta$ SL Ansatz is fixed to unity.
8

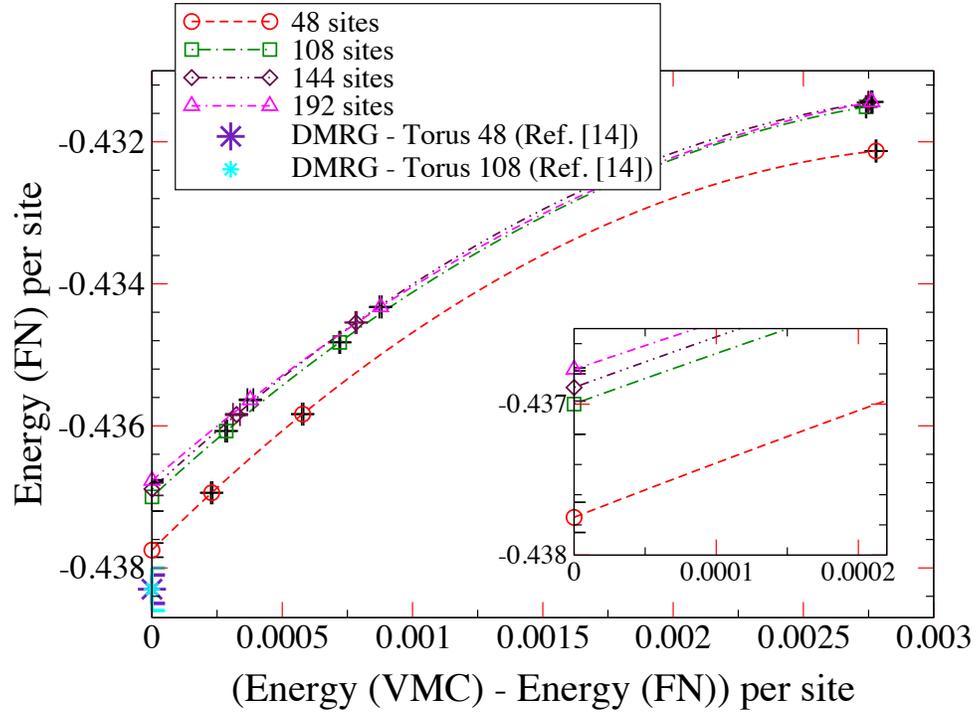

FIG. 9. (Color online) For all four cluster sizes, the fixed-node Monte Carlo energies have been extrapolated to the zero of the difference $E_{\text{VMC}} - E_{\text{FN}}$. The extrapolated values are within error bars of the zero-variance extrapolated values given in Table. I of the main paper, this shows that the VMC estimates are exact.

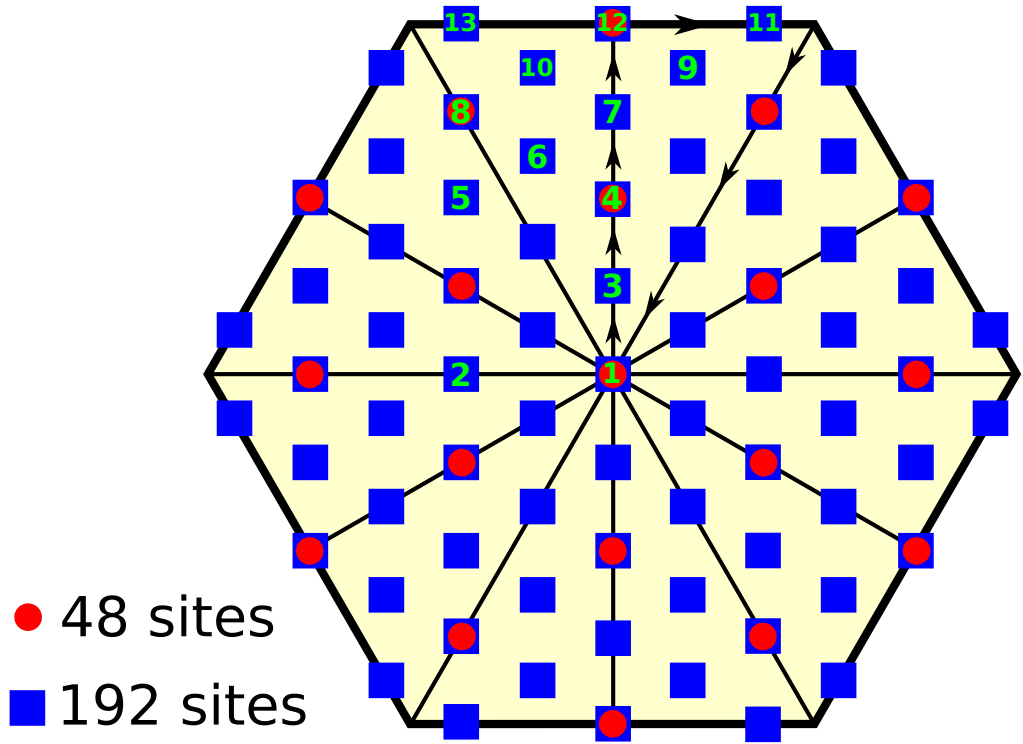

FIG. 10. (Color online) The Brillouin zone and the available crystal momenta for the 48 and the 192 site clusters are marked. Also, the 13 inequivalent momenta for the 192 site cluster are labelled with numbers.

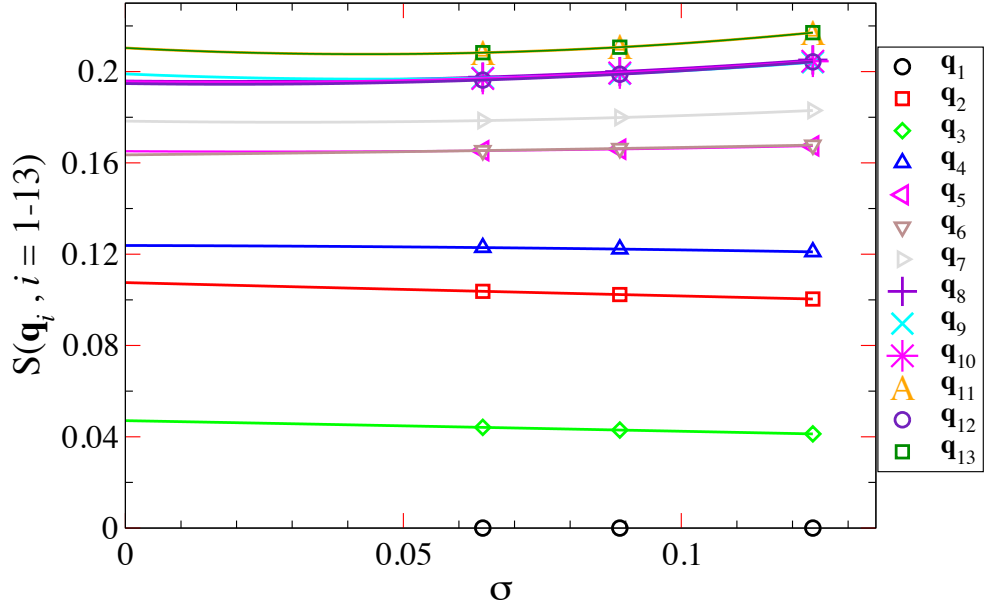

FIG. 11. (Color online) Quadratic fit of zero-$\sigma$ extrapolation of the S($\mathbf{q}$) for the 13 inequivalent momenta on the 192-site cluster. The starting wave function is the $U(1)$ Dirac spin liquid on which one and two Lanczos steps have been performed. The numbering of the momenta is according to the above given Fig. 6.

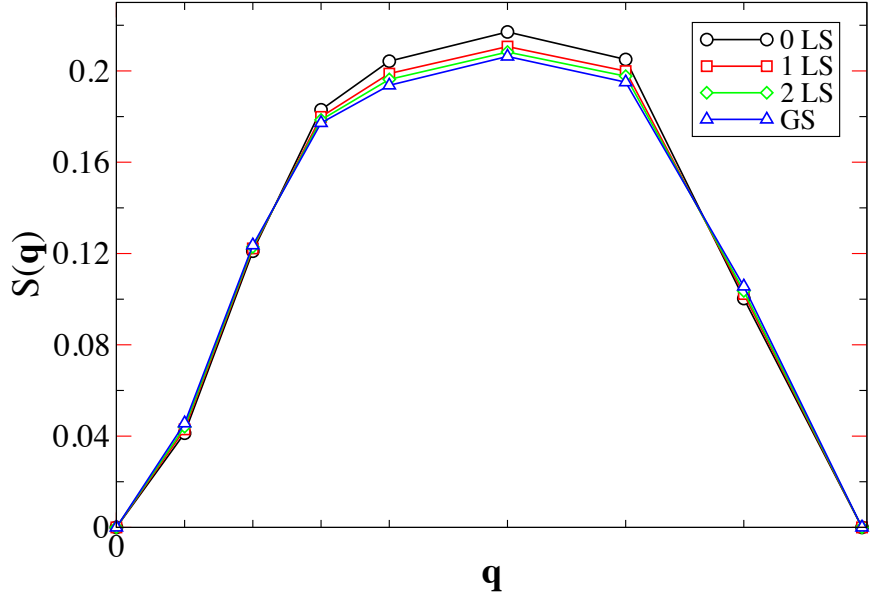

FIG. 12. (Color online) On the 192-site cluster, the S($\mathbf{q}$) are plotted for the $\mathbf{q}$ points lying on the path marked with arrows in the Brillouin zone, namely for the $U(1)$ Dirac spin liquid and the corresponding wave functions obtained by applying one, and two Lanczos steps on it, and also the extrapolated ground state. The error bars of S($\mathbf{q}$) are of the order of $10^{-4}$.




\begin{thebibliography}{99}

\bibitem{Anderson-1973} P.~W.~Anderson, \href{http://dx.doi.org/10.1016/0025-5408(73)90167-0}{Mater.~Res.~Bull.~{\bf 8}, 153 (1973)}.
\bibitem{Anderson-1987} P.~W.~Anderson, \href{http://dx.doi.org/10.1126/science.235.4793.1196}{Science.~{\bf 235}, 1196 (1987)}.
\bibitem{Balents-2010} L.~Balents, \href{http://dx.doi.org/10.1038/nature08917}{Nature (London)~{\bf 464}, 199 (2010)}.
\bibitem{Elser-1989} V.~Elser, \href{http://dx.doi.org/10.1103/PhysRevLett.62.2405}{Phys.~Rev.~Lett.~{\bf 62}, 2405 (1989)}.
\bibitem{Lecheminant-1997} P.~Lecheminant, B.~Bernu, C.~Lhuillier, L.~Pierre, and P.~Sindzingre, \href{http://dx.doi.org/10.1103/PhysRevB.56.2521}{Phys.~Rev.~B~{\bf 56}, 2521 (1997)}.
\bibitem{Sindzingre-2009} P.~Sindzingre and C.~Lhuillier, \href{http://dx.doi.org/10.1209/0295-5075/88/27009}{EPL~{\bf 88}, 27009 (2009)}.
\bibitem{Nakano-2011} H.~Nakano and T.~Sakai, \href{http://dx.doi.org/10.1143/JPSJ.80.053704}{J.~Phys.~Soc.~Jpn.~{\bf 80}, 053704 (2011)}.
\bibitem{Lauchli-2011} A.~M.~L\"auchli, J.~Sudan, and E.~S.~S\o rensen, \href{http://dx.doi.org/10.1103/PhysRevB.83.212401}{Phys.~Rev.~B~{\bf 83}, 212401 (2011)}.
\bibitem{Singh-1992} R.~R.~P.~Singh and D.~A.~Huse, \href{http://dx.doi.org/10.1103/PhysRevLett.68.1766}{Phys.~Rev.~Lett.~{\bf 68}, 1766 (1992)}.
\bibitem{Misguich-2005} G.~Misguich and B.~Bernu, \href{http://dx.doi.org/10.1103/PhysRevB.71.014417}{Phys.~Rev.~B~{\bf 71}, 014417 (2005)}.
\bibitem{Dang-2011} L.~Dang, S.~Inglis, and R.~G.~Melko, \href{http://dx.doi.org/10.1103/PhysRevB.84.132409}{Phys.~Rev.~B~{\bf 84}, 132409 (2011)}.
\bibitem{Balents-2002} L.~Balents, M.~P.~A.~Fisher, and S.~M.~Girvin, \href{http://dx.doi.org/10.1103/PhysRevB.65.224412}{Phys.~Rev.~B~{\bf 65}, 224412 (2002)}.
\bibitem{Yan-2011} S.~Yan, D.~A.~Huse, and S.~R.~White, \href{http://dx.doi.org/10.1126/science.1201080}{Science.~{\bf 332}, 1173 (2011)}.
\bibitem{Depenbrock-2012} S.~Depenbrock, I.~P.~McCulloch, and U.~Schollw\"ock, \href{http://dx.doi.org/10.1103/PhysRevLett.109.067201}{Phys.~Rev.~Lett.~{\bf 109}, 067201 (2012)}.
\bibitem{Ran-2007} Y.~Ran, M.~Hermele, P.~A.~Lee, and X.~G.~Wen, \href{http://dx.doi.org/10.1103/PhysRevLett.98.117205}{Phys.~Rev.~Lett.~{\bf 98}, 117205 (2007)}.
\bibitem{Hermele-2008} M.~Hermele, Y.~Ran, P.~A.~Lee, and X.~G.~Wen, \href{http://dx.doi.org/10.1103/PhysRevB.77.224413}{Phys.~Rev.~B~{\bf 77}, 224413 (2008)}.
\bibitem{Ma-2008} O.~Ma and J.~B.~Marston, \href{http://dx.doi.org/10.1103/PhysRevLett.101.027204}{Phys.~Rev.~Lett.~{\bf 101}, 027204 (2008)}.
\bibitem{Iqbal-2011a} Y.~Iqbal, F.~Becca, and D.~Poilblanc, \href{http://dx.doi.org/10.1103/PhysRevB.83.100404}{Phys.~Rev.~B~{\bf 83}, 100404(R) (2011)}.
\bibitem{Iqbal-2011b} Y.~Iqbal, F.~Becca, and D.~Poilblanc, \href{http://dx.doi.org/10.1103/PhysRevB.84.020407}{Phys.~Rev.~B~{\bf 84}, 020407(R) (2011)}.
\bibitem{Iqbal-2012} Y.~Iqbal, F.~Becca, and D.~Poilblanc, \href{http://dx.doi.org/10.1088/1367-2630/14/11/115031}{New~J.~Phys.~{\bf 14}, 115031 (2012)}.
\bibitem{Marston-1991} J.~B.~Marston and C.~Zeng, \href{http://dx.doi.org/10.1063/1.347830}{J.~Appl.~Phys.~{\bf 69}, 5962 (1991)}.
\bibitem{Nikolic-2003} P.~Nikolic and T.~Senthil, \href{http://dx.doi.org/10.1103/PhysRevB.68.214415}{Phys.~Rev.~B~{\bf 68}, 214415 (2003)}.
\bibitem{Hwang-2011} K.~Hwang, Y.~B.~Kim, J.~Yu, and K.~Park, \href{http://dx.doi.org/10.1103/PhysRevB.84.205133}{Phys.~Rev.~B~{\bf 84}, 205133 (2011)}.
\bibitem{Zeng-1995} C.~Zeng and V.~Elser, \href{http://dx.doi.org/10.1103/PhysRevB.51.8318}{Phys.~Rev.~B~{\bf 51}, 8318 (1995)}.
\bibitem{Poilblanc-2010} D.~Poilblanc, M.~Mambrini, and D.~Schwandt, \href{http://dx.doi.org/10.1103/PhysRevB.81.180402}{Phys.~Rev.~B~{\bf 81}, 180402(R) (2010)}.
\bibitem{Schwandt-2010} D.~Schwandt, M.~Mambrini, and D.~Poilblanc, \href{http://dx.doi.org/10.1103/PhysRevB.81.214413}{Phys.~Rev.~B~{\bf 81}, 214413 (2010)}.
\bibitem{Ralko-2010} A.~Ralko and D.~Poilblanc, \href{http://dx.doi.org/10.1103/PhysRevB.82.174424}{Phys.~Rev.~B~{\bf 82}, 174424 (2010)}.
\bibitem{Poilblanc-2011} D.~Poilblanc and G.~Misguich, \href{http://dx.doi.org/10.1103/PhysRevB.84.214401}{Phys.~Rev.~B~{\bf 84}, 214401 (2011)}.
\bibitem{Singh-2007} R.~R.~P.~Singh and D.~A.~Huse, \href{http://dx.doi.org/10.1103/PhysRevB.76.180407}{Phys.~Rev.~B~{\bf 76}, 180407(R) (2007)}.
\bibitem{Singh-2008} R.~R.~P.~Singh and D.~A.~Huse, \href{http://dx.doi.org/10.1103/PhysRevB.77.144415}{Phys.~Rev.~B~{\bf 77}, 144415 (2008)}.
\bibitem{Evenbly-2010} G.~Evenbly and G.~Vidal, \href{http://dx.doi.org/10.1103/PhysRevLett.104.187203}{Phys.~Rev.~Lett.~{\bf 104}, 187203 (2010)}.
\bibitem{Gotze-2011} O.~G\"otze, D.~J.~J.~Farnell, R.~F.~Bishop, P.~H.~Y.~Li, and J.~Richter, \href{http://dx.doi.org/10.1103/PhysRevB.84.224428}{Phys.~Rev.~B~{\bf 84}, 224428 (2011)}.
\bibitem{Sachdev-1992} S.~Sachdev, \href{http://dx.doi.org/10.1103/PhysRevB.45.12377}{Phys.~Rev.~B~{\bf 45}, 12377 (1992)}.
\bibitem{Wen-2002} X.~G.~Wen, \href{http://dx.doi.org/10.1103/PhysRevB.65.165113}{Phys.~Rev.~B~{\bf 65}, 165113 (2002)}.
\bibitem{Misguich-2002} G.~Misguich, D.~Serban, and V.~Pasquier, \href{http://dx.doi.org/10.1103/PhysRevLett.89.137202}{Phys.~Rev.~Lett.~{\bf 89}, 137202 (2002)}.
\bibitem{Hermele-2004} M.~Hermele, T.~Senthil, M.~P.~A.~Fisher, P.~A.~Lee, N.~Nagaosa, and X.~G.~Wen, \href{http://dx.doi.org/10.1103/PhysRevB.70.214437}{Phys.~Rev.~B~{\bf 70}, 214437 (2004)}.
\bibitem{Yang-2012} F.~Yang and H.~Yao, \href{http://dx.doi.org/10.1103/PhysRevLett.109.147209}{Phys.~Rev.~Lett.~{\bf 109}, 147209 (2012)}.
\bibitem{Shores-2005} M.~P.~Shores, E.~A.~Nytko, B.~M.~Bartlett, and D.~G.~Nocera, \href{http://dx.doi.org/10.1021/ja053891p}{J.~Am.~Chem.~Soc.~{\bf 127}, 13462 (2005)}.
\bibitem{Ofer-2006} O.~Ofer, A.~Keren, E.~A.~Nytko, M.~P.~Shores, B.~M.~Bartlett, D.~G.~Nocera, C.~Baines, and A.~Amato, \href{http://arxiv.org/abs/arXiv:cond-mat/0610540}{arXiv:~0610540 (2006) [cond.mat]}.
\bibitem{Bert-2007} F.~Bert, S.~Nakamae, F.~Ladieu, D.~L'H$\hat{\rm o}$te, P.~Bonville, F.~Duc, J.-C.~Trombe, and P.~Mendels, \href{http://dx.doi.org/10.1103/PhysRevB.76.132411}{Phys.~Rev.~B.~{\bf 76}, 132411 (2007)}.
\bibitem{Lee-2007} S.~H.~Lee, H.~Kikuchi, Y.~Qiu, B.~Lake, Q.~Huang, K.~Habicht, and K.~Kiefer, \href{http://dx.doi.org/10.1038/nmat1986}{Nat.~Mater.~{\bf 6}, 853 (2007)}.
\bibitem{Lee-2008} P.~A.~Lee, \href{http://dx.doi.org/10.1126/science.1163196}{Science~{\bf 321}, 1306 (2008)}.
\bibitem{Vries-2008} M.~A.~de Vries, K.~V.~Kamenev, W.~A.~Kockelmann, J.~Sanchez-Benitez, and A.~Harrison, \href{http://dx.doi.org/10.1103/PhysRevLett.100.157205}{Phys.~Rev.~Lett.~{\bf 100}, 157205 (2008)}.
\bibitem{Imai-2008} T.~Imai, E.~A.~Nytko, B.~M.~Bartlett, M.~P.~Shores, and D.~G.~Nocera, \href{http://dx.doi.org/10.1103/PhysRevLett.100.077203}{Phys.~Rev.~Lett.~{\bf 100}, 077203 (2008)}.
\bibitem{Olariu-2008} A.~Olariu, P.~Mendels, F.~Bert, F.~Duc, J.~C.~Trombe, M.~A.~de Vries, and A.~Harrison, \href{http://dx.doi.org/10.1103/PhysRevLett.100.087202}{Phys.~Rev.~Lett.~{\bf 100}, 087202 (2008)}.
\bibitem{Mendels-2010} P.~Mendels and F.~Bert, \href{http://dx.doi.org/10.1143/JPSJ.79.011001}{J.~Phys.~Soc.~Jpn.~{\bf 79}, 011001 (2010)}.
\bibitem{Han-2011} T.~H.~Han, J.~S.~Helton, S.~Chu, A.~Prodi, D.~K.~Singh, C.~Mazzoli, P.~M\"uller, D.~G.~Nocera, and Y.~S.~Lee, \href{http://dx.doi.org/10.1103/PhysRevB.83.100402}{Phys.~Rev.~B~{\bf 83}, 100402(R) (2011)}.
\bibitem{Mendels-2007} P.~Mendels, F.~Bert, M.~A.~de Vries, A.~Olariu, A.~Harrison, F.~Duc, J.~C.~Trombe, J.~S.~Lord, A.~Amato, and C.~Baines, \href{http://dx.doi.org/10.1103/PhysRevLett.98.077204}{Phys.~Rev.~Lett.~{\bf 98}, 077204 (2007)}.
\bibitem{Helton-2007} J.~S.~Helton, K.~Matan, M.~P.~Shores, E.~A.~Nytko, B.~M.~Bartlett, Y.~Yoshida, Y.~Takano, A.~Suslov, Y.~Qiu, J.~H.~Chung, D.~G.~Nocera, and Y.~S.~Lee, \href{http://dx.doi.org/10.1103/PhysRevLett.98.107204}{Phys.~Rev.~Lett.~{\bf 98}, 107204 (2007)}.
\bibitem{Kermarrec-2011} E.~Kermarrec, P.~Mendels, F.~Bert, R.~H.~Colman, A.~S.~Wills, P.~Strobel, P.~Bonville, A.~Hillier, and A.~Amato, \href{http://dx.doi.org/10.1103/PhysRevB.84.100401}{Phys.~Rev.~B~{\bf 84}, 100401 (2011)}.
\bibitem{Wulferding-2010} D.~Wulferding, P.~Lemmens, P.~Scheib, J.~R\"oder, P.~Mendels, S.~Chu, T.~Han, and Y.~S.~Lee, \href{http://dx.doi.org/10.1103/PhysRevB.82.144412}{Phys.~Rev.~B~{\bf 82}, 144412 (2010)}.
\bibitem{Sorella-2001} S.~Sorella, \href{http://dx.doi.org/10.1103/PhysRevB.64.024512}{Phys.~Rev.~B~{\bf 64}, 024512 (2001)}.
\bibitem{Haaf-1995} D.~F.~B.~ten Haaf, H.~J.~M.~van Bemmel, J.~M.~J.~van Leeuwen, W.~van Saarloos, and D.~M.~Ceperley, \href{http://dx.doi.org/10.1103/PhysRevB.51.13039}{Phys.~Rev.~B~{\bf 51}, 13039 (1995)}.
\bibitem{Wen-1991} X.~G.~Wen, \href{http://dx.doi.org/10.1103/PhysRevB.44.2664}{Phys.~Rev.~B~{\bf 44}, 2664 (1991)}.
\bibitem{Sorella-2005} S.~Sorella, \href{http://dx.doi.org/10.1103/PhysRevB.71.241103}{Phys.~Rev.~B~{\bf 71}, 241103 (2005)}.
\bibitem{Yunoki-2006} S.~Yunoki and S.~Sorella, \href{http://dx.doi.org/10.1103/PhysRevB.74.014408}{Phys.~Rev.~B~{\bf 74}, 014408 (2006)}.
\bibitem{Lu-2011} Y.~M.~Lu, Y.~Ran, and P.~A.~Lee, \href{http://dx.doi.org/10.1103/PhysRevB.83.224413}{Phys.~Rev.~B~{\bf 83}, 224413 (2011)}.
\bibitem{Jiang-2008} H.~C.~Jiang, Z.~Y.~Weng, and D.~N.~Sheng, \href{http://dx.doi.org/10.1103/PhysRevLett.101.117203}{Phys.~Rev.~Lett.~{\bf 101}, 117203 (2008)}.
\bibitem{suppmat} See Supplemental Material at the end of the main paper for a figure of the kagome lattice (Fig.~5), additional ground state energy zero-variance extrapolations data for all clusters (Figs.~6, 7, and 8), fixed-node ground state energy extrapolation vs the difference $E_{\rm VMC} - E_{\rm FN}$ for all clusters (Fig.~9), available crystal momenta of the $48$- and $192$-site clusters (Fig.~10), zero-variance extrapolation of static spin structure factors (Fig. 11), and the structure factors along a path in the Brillouin zone (Fig.~12).
\bibitem{privateDMRG} On the $48$-site torus, $E/J=-0.4366$ corresponds to $m=3000$ states in the $SU(2)$ implementation of DMRG (U. Schollw\"ock, private communication). Recent unpublished calculations by S. R. White using the $U(1)$ implementation of DMRG  give an energy per site of $E/J=-0.4381$ for $m=19860$ states (private communication). On the $108$-site torus, $E/J=-0.4316$ corresponds to $m=6000$ states in the $U(1)$ DMRG of Ref.~\onlinecite{Jiang-2008}. The $SU(2)$ DMRG with $m=1400$ states gives $E/J=-0.4285$ (U. Schollw\"ock, private communication). The best extrapolated ground-state energy estimate from DMRG on the $48$-site torus is $E/J=-0.4383(2)$ and on the $108$-site torus is $E/J=-0.4383(3)$ (see Ref.~\onlinecite{Depenbrock-2012}).
\bibitem{Motrunich-2011a} T.~Tay and O.~I.~Motrunich, \href{http://dx.doi.org/10.1103/PhysRevB.84.020404}{Phys.~Rev.~B~{\bf 84}, 020404(R) (2011)}.
\bibitem{Clark-2012} B.~K.~Clark, J.~M.~Kinder, E.~Neuscamman, G.~K.-L.~Chan, and M.~J.~Lawler, \href{http://arxiv.org/abs/1210.1585}{arXiv:~1210.1585 (2012) [cond.mat]}.
\bibitem{Capponi-2012} S.~Capponi, V.~Ravi~Chandra, A.~Auerbach, and M.~Weinstein, \href{http://arxiv.org/abs/1210.5519}{arXiv:~1210.5519 (2012) [cond.mat]}.
\bibitem{Motrunich-2011} T.~Tay and O.~I.~Motrunich, \href{http://dx.doi.org/10.1103/PhysRevB.84.193102}{Phys.~Rev.~B~{\bf 84}, 193102 (2011)}.
\bibitem{Yang-1993} K.~Yang, L.~K.~Warman, and S.~M.~Girvin, \href{http://dx.doi.org/10.1103/PhysRevLett.70.2641}{Phys.~Rev.~Lett.~{\bf 70}, 2641 (1993)}.
\bibitem{Messio-2011} L.~Messio, B.~Bernu, and C.~Lhuillier, \href{http://dx.doi.org/10.1103/PhysRevLett.108.207204}{Phys.~Rev.~Lett.~{\bf 108}, 207204 (2012)}.

\end{thebibliography}
\end{document}